\documentclass[conference]{IEEEtran}

\usepackage[utf8]{inputenc}
\usepackage{amsmath, amsthm}
\usepackage{subfigure}
\usepackage{graphicx}

\usepackage{listings}
\usepackage{paralist}

\usepackage{url}
\usepackage{dsfont}
\usepackage{xspace}
\usepackage{array}

\usepackage[inline]{enumitem} 

\usepackage{todonotes}

\setuptodonotes{fancyline, color=blue!30}
\presetkeys{todonotes}{inline}{}

\usepackage{algorithm,etoolbox}
\usepackage{algpseudocode}

\usepackage{hyperref}

\usepackage{booktabs}
\usepackage{pgf,tikz}
\usepackage{forest}
\usepackage{varwidth}
\usepackage{marvosym}
\usetikzlibrary{shapes}
\usetikzlibrary{arrows,matrix}
\usetikzlibrary{positioning}
\usetikzlibrary{arrows,automata,calc,positioning}
\usetikzlibrary{shapes.arrows,decorations.pathreplacing,backgrounds}
\usetikzlibrary{patterns}
\usetikzlibrary{shapes,arrows,backgrounds,positioning,calc, automata, shadows, backgrounds,fit, arrows,graphs, trees, decorations.pathreplacing, patterns}
\tikzstyle{component}=[draw, text centered, text width=5em]
\tikzstyle{arbitrer}=[draw, circle, text centered, text width=1em]
\tikzstyle{ann} = [above, text width=5em]

{\itshape}

\newtheorem{hybrid-case}{Category}

\newtheoremstyle{example}
  {\topsep}
  {\topsep}
  {\itshape}
  {}
  {\bfseries}
  {.}
  {.5em}
  {}

\theoremstyle{example}
\newtheorem{example}{Example}

\newcommand{\phylogun}{{\sc Phylog}\xspace}
\newcommand{\phylog}{{\sc Phylog}\xspace}

\newcommand{\multicore}{multi-core\xspace}

\newcommand{\mcp}{\multicore processor\xspace}
\newcommand{\mcps}{\multicore processors\xspace}

\newcommand{\armx}{{\sc arm}\xspace}

\newcommand{\sm}{SM\xspace}
\newcommand{\gpu}{GPU\xspace}
\newcommand{\gpgpu}{GPGPU\xspace}

\newcommand{\gpus}{GPUs\xspace}
\newcommand{\soc}{SoC\xspace}
\newcommand{\asic}{{\sc Asic}\xspace}

\newcommand{\cots}{COTS\xspace}
\newcommand{\fpga}{FPGA\xspace}
\newcommand{\simd}{SIMD\xspace}
\newcommand{\simt}{SIMT\xspace}

\newcommand{\valu}{VAU\xspace}

\newcommand{\PHAC}{PHAC\xspace}

\newcommand{\keystone}{\textsc{Keystone}\xspace}

\newcommand{\pml}{\textsc{PML}\xspace}

\newcommand{\edHW}{ED-80/DO-254\xspace}
\newcommand{\amcMCP}{AMC20-193\xspace}
\newcommand{\amcCOTS}{AMC20-152A\xspace}

\AtBeginEnvironment{code}{\vspace*{0.3cm}\noindent\hspace{-.3cm}\begin{minipage}{1.07\linewidth}}
\AtEndEnvironment{code}{\end{minipage}}

\lstdefinelanguage{idp}
{
  language=C,
  morekeywords={!, ?}
}

{ \begin{list}%
	{$-$}%
	{\setlength{\labelsep}{5pt}
	\setlength{\leftmargin}{25pt}
	\setlength{\labelwidth}{0pt}
	\setlength{\listparindent}{\listparindent}
	\setlength{\itemindent}{\listparindent}
	\setlength{\itemsep}{-2pt}
        \setlength{\topsep}{-2pt}
	}}%
{ \end{list} }

\begin{document}

\title{Towards the Certification of Hybrid Architectures: Analysing Interference on Hardware Accelerators through PML}

\ifdefined\anonymous
\else
\author{
  \IEEEauthorblockN{Benjamin Lesage$^1$, Frederic Boniol$^1$, Kevin Delmas$^1$,  Adrien Gauffriau$^2$, Alfonso Mascarenas-Gonzalez$^1$,  Claire Pagetti$^1$}
  \IEEEauthorblockA{$^1$ ONERA, Toulouse, France, $^2$ Airbus, Toulouse, France}
    }
\fi

\maketitle

\begin{abstract}


  The emergence of Deep Neural Network (DNN) and machine learning-based applications paved the way for a new generation of hybrid hardware platforms. Hybrid platforms embed several cores and accelerators in a small package. 
  However, in order to satisfy the Size, Weight and Power (SWaP) constraints,
  limited and shared resources are integrated.
This paper presents an overview of the standards applicable to the certification of hybrid platforms and an early mapping of their objectives to said platforms. In particular, we consider how the classification of \amcCOTS for airborne electronic hardware applies to hybrid platforms. We also consider \amcMCP for multi-core platforms, and how this standard fits different types of accelerators.


\end{abstract}

\IEEEpeerreviewmaketitle

\maketitle



\section{Introduction}
\label{sec:introduction}

New software paradigms and capabilities drive the demand for additional computing power in avionic systems.
Hybrid architectures can, in a small SWaP package, support this demand.
They embed on the same platform general-purpose cores, and specialised accelerators which can support some of the additional workload.
However, like any other hardware platform, they need to go through a stringent certification process before they are deployed in avionic system.

The European Union Aviation Safety Agency (EASA) and Federal Aviation Administration (FAA) respectively define Acceptable Means of Compliance (AMC) and Advisory Circulars (AC), setting down objectives applicants to the certification process satisfy.
The joint A(M)C \amcCOTS and \amcMCP in particular define objectives for the respective certification of hardware platforms and multi-core processors.

The \phylog methodology \cite{erts20} was proposed as mean of supporting applicants, especially regarding \amcMCP on \mcps. 
\phylog is based on the definition of argumentation patterns for the certification objectives in \amcMCP, with each objective decomposed in supporting claims, strategies, or evidences.
At the core of the methodology, the \phylog Modelling Language (\pml) \cite{phylog_erts22}
captures knowledge about a platform, both hardware and software aspects, and their configuration.
\pml supports analyses to fulfil claims in the certification patterns instantiated for the platform.

The contributions of this paper are to 
present an overview of the objectives applicable to hybrid platforms. We also identify the issues related to modelling the accelerators in such platforms and propose related \pml model templates.
This paper is organised as follows. 
Section~\ref{sec:phylog} briefly recaps the \phylog methodology, with Section~\ref{sec:pml} providing an introduction to \pml.
An example of accelerator and its hybrid platform is introduced in Section~\ref{sec:platform} to support further discussions and examples.
In the context of hybrid platforms, we identified two relevant AMC: \amcCOTS~\cite{amc20152a} and \amcMCP~\cite{amc20193} discussed respectively in Section~\ref{sec:amc cots} and Section~\ref{sec:amc mcp}.
Section~\ref{sec:related work} briefly discusses related work, before
 Section~\ref{sec:conclusion} recaps the discussion and outlines perspectives.


\section{PHYLOG Methodology}
\label{sec:phylog}

\begin{figure}[ht!]
    \centering
    \includegraphics[width=0.60\linewidth]{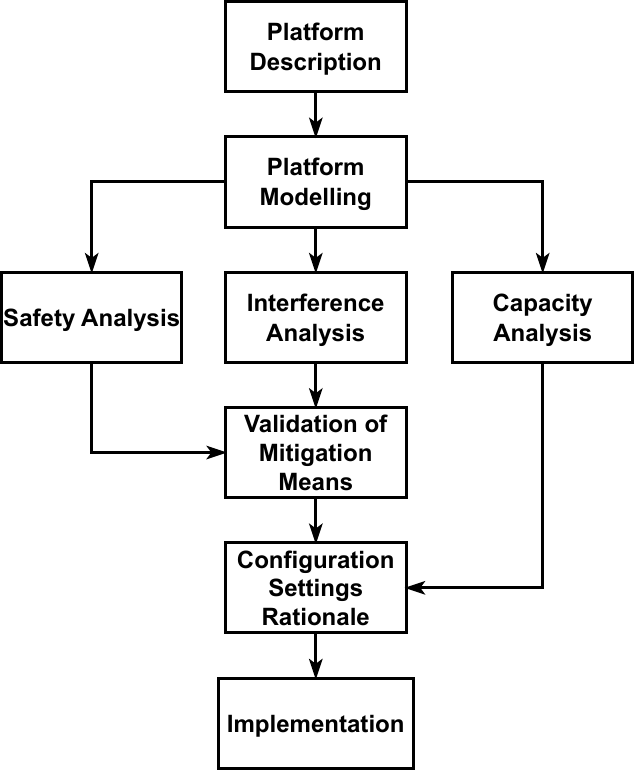}
    \caption{Overview of \phylogun methodology}
    \label{fig:phylog-methodology-overview}
\end{figure}

The \phylogun methodology~\cite{erts20} describes the activities to produce the elements for instantiating the \phylogun argumentation patterns.  
These patterns were derived from the objectives defined in \amcMCP, on \mcps, to build an argumentation strategy for certification.
They decompose the top-level AMC objectives into supporting claims, strategies, evidence, and warrants.
An overview of the methodology is presented in Figure~\ref{fig:phylog-methodology-overview}. It is composed of eight main activities:
\begin{itemize}
\item \textbf{Platform description} captures the knowledge about the platform characteristics based on the available documents and the applicant's assessments. It also captures the target configuration, including hardware and software settings such as the mapping of applications hosted on the platform to cores. 
\item \textbf{Platform modelling} formalises the platform description knowledge in order to support further analyses. It is based on \pml.
While not an objective of \amcMCP, it allows running the supporting automatic safety, capacity and interference analyses in order to contribute to said objectives.
\item  \textbf{Safety analysis} identifies and evaluates the failures and alterations which can affect the platform and hosted applications.
\item \textbf{Interference analysis} enables the identification of interferences via interference calculus and the classification of their effects.
\item \textbf{Capacity analysis} enables the verification of shared resources' usage, ensuring the demand for resources of the platform never exceeds their capacity. 
\item \textbf{Validation of mitigation means} encompasses the design and validation of mitigation means for failure, interference, and other alterations identified in earlier activities.
\item \textbf{Configuration settings rationale} justifies that all configuration settings support the requirements on the platform, or are harmless to them.
\item \textbf{Implementation} concerns the certification of the system implementation on the platform. It is associated with the DO-178C standard and out of the context of \phylogun.
\end{itemize}

Note that the activities form an inherently iterative process. As an example, the interference analysis may highlight a misunderstood interference channel, feeding back into the platform description and its model. 

We focus in the following on the platform aspects (description and modelling), as they are the most relevant to hybrid platforms. We consider specifically the use of \pml, and its limitations, to model accelerators. 
The use of \pml would thus allow for the application of existing \phylog-based analyses~\cite{erts20}, discussed in other work for interference or safety, to instantiate the \phylog certification patterns for hybrid platforms. 
\pml is introduced in the next section.

\section{PML}
\label{sec:pml}

\pml, the \phylog Modelling Language \cite{phylog_erts22}, is a Domain Specific Language embedded with the SCALA language to capture the description of a platform.
A hardware platform is modelled in \pml as a collection of components, capturing the functional blocks of a \mcp, e.g. a core, cache, memory, or bus, and links between components.
Composite components encapsulate one or more components, composite or atomic, to allow for the hierarchical specification of a model.
\emph{Atomic} components provide generic services to the software hosted by the platform, such as a \emph{load} from the main memory or a \emph{store} to a configuration register.

The relationship between a component and other services of the platform defines its role in the model.
\emph{Initiator} components, such as a core, call services from other components on the platform, most often as a result of software running on the initiator, be it a user application or platform-embedded micro-code.
\emph{Target} components, such as the main memory, expose services to satisfy transactions from other components.
\emph{Transporter} components, such as an interconnect, process transactions between an initiator and its target.

A \emph{transaction} is a footprint of a use of the platform by a software component.
A transaction more formally captures the set of components, and their services, used by a request from an initiator to a target.
A transaction must follow a valid path in the platform, through the links between its components.
Services thus model the dependencies between the software and the hardware.

\begin{example}
\label{ex-hybrid-keystone}

To exemplify the use of PML, we consider a representation of the \keystone TCI6630K2L from Texas Instruments. An overview of the \keystone is presented in Figure~\ref{fig:ex-keystone}.
It is composed of a four C66 DSP pack where cores are characterised by dedicated L1 and L2 caches, and a memory extension and protection unit (MPAX). The platform also comprises a 2 \armx A15 pack where cores are characterised by dedicated L1 caches, memory management units (MMU), and a shared L2. In addition, it includes a central memory system giving access to SRAM and external DDR. Memory accesses are managed by a Multicore Shared Memory Controller (MSMC). A set of I/O and utility peripherals (e.g. GPIO, UART, boot) is also present on the platform and an ultra speed bus (TeraNet) connects the peripherals, the memories, and the cores altogether.

\begin{figure}[!ht]
        \centering
		\includegraphics[scale=0.40]{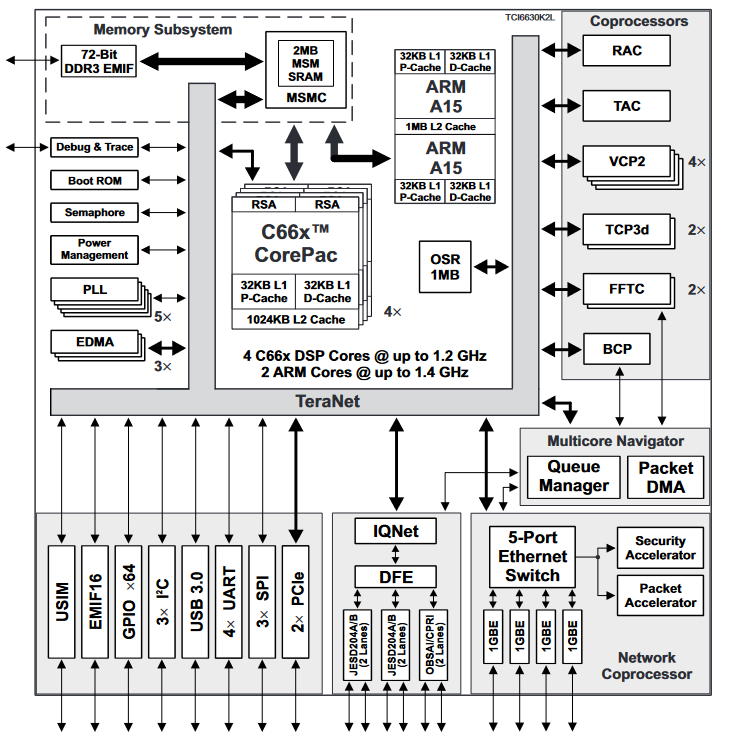}
	    \caption{Overview of the TI \keystone TCI6630K2L}
	    \label{fig:ex-keystone}
\end{figure}


Figure~\ref{fig:ex-keystone-pml} illustrates a PML model for a simplified version of the \keystone\footnote{
For the sake of brevity, coprocessors have been omitted, 
as well as implicit links between stacked components.
Peripherals have been simply classified as targets.
}. 
This basic model includes:
\begin{itemize}
\item Cores as initiators: 4 C66 DSP, and 2 \armx A15 cores;
\item Memories as targets: DDR, SRAM, and all caches;
\item Peripherals as targets: GPIO, I2C, SPI port, PCIe, etc.;
\item Buses and Memory protection units as transporters: the TeraNet bus connected to the Memory Shared Multicore Controller (MSMC), memory and cache controllers, etc.
\end{itemize}

\begin{figure}[!ht]
	     \centering
	     \includegraphics[scale=0.70]{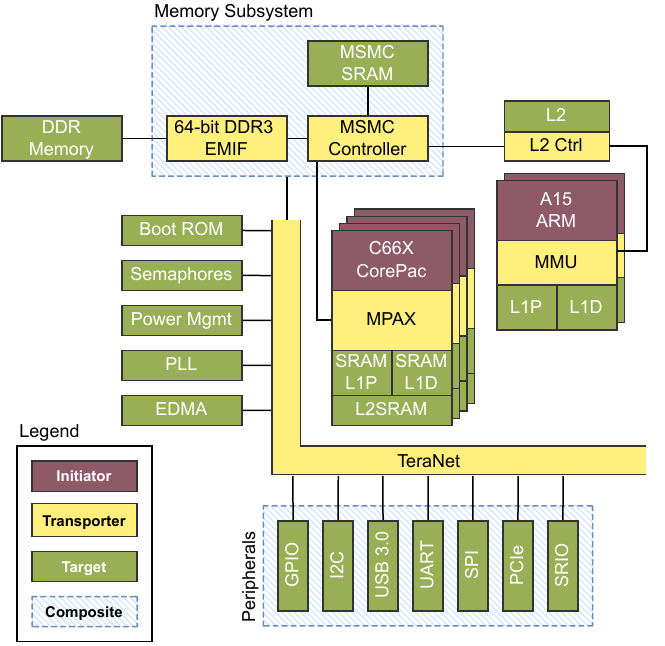}
	     \caption{Simplified PML model for the \keystone platform}
		\label{fig:ex-keystone-pml}
\end{figure}

\end{example}





\section{Hybrid architectures - The GPU Example}
\label{sec:platform}

To support the discussion around hybrid platforms, we introduce an example of accelerator: Graphical Processing Units (\gpu).
Compared to traditional CPUs, \gpus feature numerous cores with simpler control flow but efficient data ones. \gpu cores tend to work in a lockstep-like fashion called Single Instruction Multiple Threads (\simt) in reference to \simd (Single Instruction Multiple Data). Internal scheduling policies on the \gpu aim to maximise core occupancy and throughput. 
With their focus on high-throughput floating point computation, \gpu are well suited to the acceleration of neural network workloads. Their reuse has been facilitated by the advent of General Purpose \gpu programming frameworks (\gpgpu). 


There has been considerable effort to characterise the behaviour of GPU accelerators, in particular work on NVIDIA \gpu \cite{charac:Volta, PastisWCETGPU, charac:Nathan, charac:cuda_sched} and the assorted \gpgpu CUDA software stack \cite{charac:pitfalls, charac:Nathan, charac:Tanya, charac:cuda_sched}. 
These efforts highlight the difficulty of characterising complex, multi-core, \cots (Commercially available Off-the-Shelf) platforms. To the best of our knowledge, PasTiS~\cite{PastisWCETGPU} and the hybrid analysis in~\cite{HybridWCETGPU} are some of the few efforts to build a GPU model respectively for static and hybrid WCET analysis. The inherent parallelism at the application-level, as opposed to instruction-level like vectorised arithmetic units~\cite{10.1145/3561054, acetone_simd}, can also pose problems for WCET and interference analyses~\cite{LisperWCET2012}.

\begin{example}
\label{ex-hybrid-volta}
\it
The NVIDIA Jetson AGX Xavier~\cite{NVIDIA_Xavier_TRM} is a high-performance \soc designed for embedded systems. The Xavier uses an 8-core ``Carmel" \armx processor, organised in clusters of 2 cores. The ``Carmel" processor complies the \armx v8.2A specification, but it is unclear if it is based off an existing \armx design (e.g. the Cortex-A78) and the level of customisation introduced by NVIDIA. The Xavier features amongst other accelerators a \gpu using the Volta architecture, highlighted in Figure~\ref{fig-hybrid-jetson}. The \gpu is composed of 512 cores, grouped in 8 Streaming Multiprocessors (SM). The Volta \gpu shares a memory fabric with other accelerators, and the memory controller with the CPU.

\begin{figure}[!htb]
    \centering
		\centering
		\includegraphics[width=0.95\linewidth]{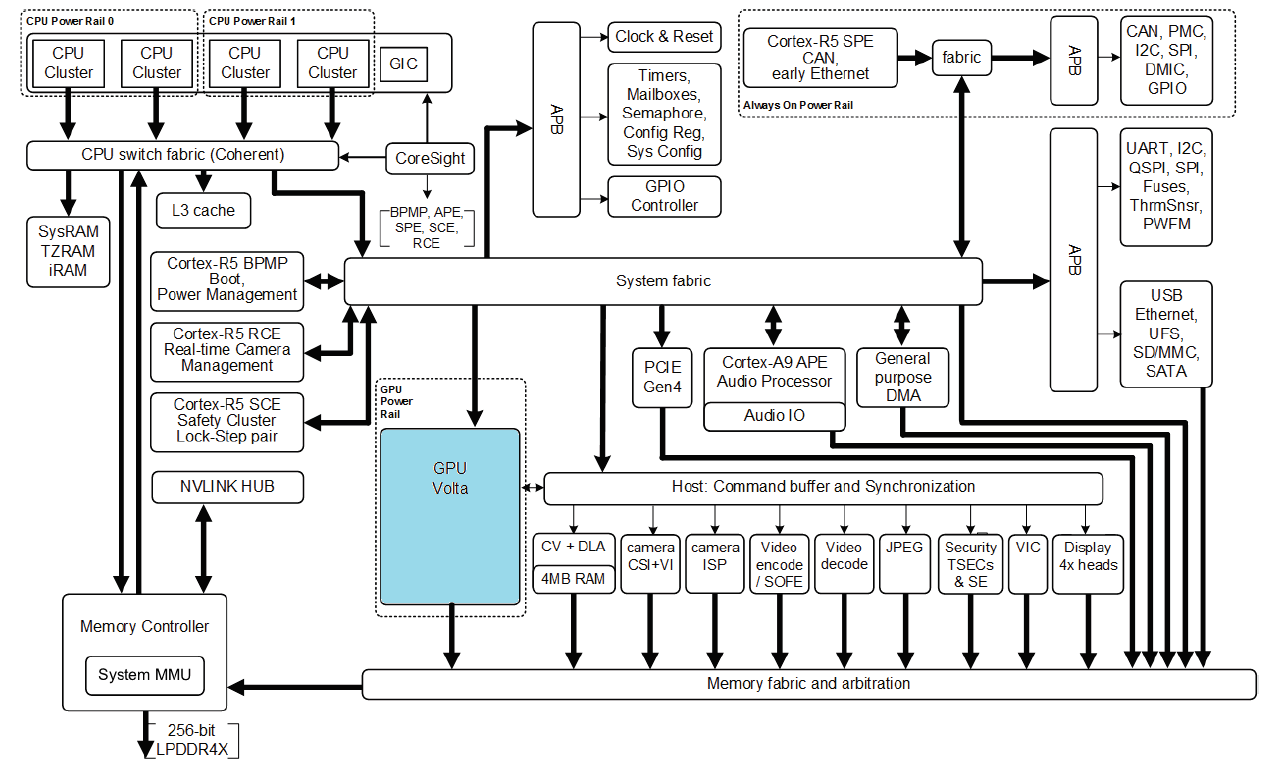}
		\caption{Overview of the NVIDIA Jetson AGX Xavier
		\label{fig-hybrid-jetson}}
\end{figure}

We present in Figure~\ref{fig-hybrid-jetson-pml} a high-level \pml model of the NVIDIA AGX Xavier \soc. Fabrics and backbones act as transporters for the components of the system. The main memory is a target shared by the CPU and the GPU. The cores of the ``Carmel'' \armx processor  act as multiple initiators. As for the \keystone, we currently omit coprocessors and peripherals from the classification. A key question is: \emph{How to model a complex accelerator like the Volta \gpu?} 
It acts 
as an initiator, causing interference on the main memory and the controller fabric, 
and as a target for commands from the CPU.

\begin{figure}[!htb]
	     \centering
	     \includegraphics[scale=0.70]{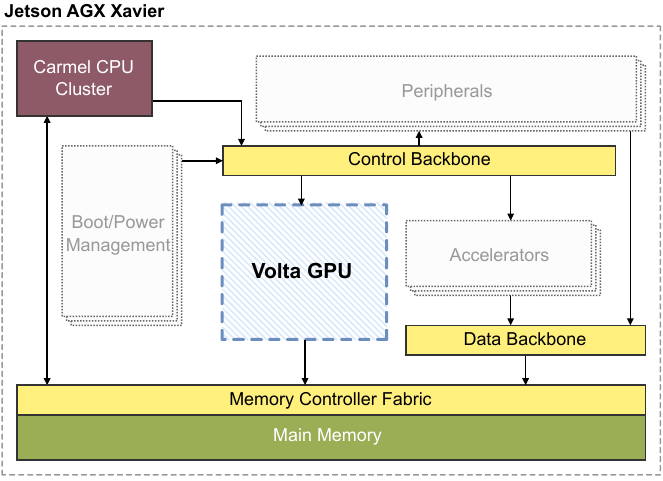}
	     \caption{Simplified PML model for the NVIDIA AGX Xavier}
		\label{fig-hybrid-jetson-pml}
\end{figure}

\end{example}


\section{\amcCOTS on hybrid architectures}
\label{sec:amc cots}

\amcCOTS discusses the certification of existing (\cots) or newly-developed platforms, the distinction between the two, and the objectives relevant to each.

\subsection{Overview of the \amcCOTS}
\label{sec:amc cots overview}


The \edHW, both dated from the year 2000, define guidance for the design of airborne electronic hardware. The \amcCOTS aims to provide additional guidance and clarification. It is thus complementary to the \amcMCP on multi-core platforms. The clarifications proposed by the \amcCOTS are important, as devices, especially \cots, become more complex and integrate in a single chip more functions than older ones. The \amcCOTS objectives are classified according to whether they apply to complex custom devices, \cots IP (design functions used to design and implement a custom device, be it a PLD, a \fpga or an \asic), or \cots devices\footnote{We omit circuit boards assemblies (CBA), as the \amcCOTS in practice redirects to \edHW.}.
Applicants for certification must address them as part of the Plan for Hardware Aspects of Certification (\PHAC), or related planning documents. 

The first distinction in the \amcCOTS is between \cots and custom functions. \cots functions (IP or devices) are, as the name implies, commercially available, off-the-shelf.  The \amcCOTS recognises the risks inherent to the use of COTS, and incomplete or incorrect documentation. \cots may not have been developed within the \edHW standard or avionic applications, nor have sufficient service experience. The development assurance for \cots items (hardware or software components \edHW) thus follows different objectives from custom devices.
Items developed and fully controlled by the applicant cannot classify as \cots.
Those items may however be previously developed hardware, which may take credit from prior deployment and in-service experience provided their new function, usage and environment conditions do not invalidate the original design assurance.

The key objectives of the process for \cots items are
\begin{enumerate*}
\item identifying used functions, and
\item assessing correct use of the \cots item.
\end{enumerate*}
The used functions need to support the system requirements on the device. Unused functions, such as unused cores on a MCP (as per \amcMCP), need to be properly deactivated, with means of mitigation to prevent their inadvertent activation. 
Correct use of a \cots item requires to assess its integration against the operation conditions, such as temperature or input parameter ranges, defined by the manufacturer. This may preclude the use of undefined or undocumented configurations, unless their reliability can be established.
The identification of failure modes\footnote{Single Event Effects (SEE) are explicitly omitted from the \amcCOTS scope.} and the item configuration also need to be considered.
This includes identifying if any microcode may contribute to a used function. Microcode is a hardware-level set of instructions, typically stored in the \cots item. It may be qualified by the manufacturer, if left unmodified, or require a separate mean of compliance.




Devices are further classified into simple or complex ones as defined by the \edHW. The classification
captures whether a comprehensive verification of the device is realistic. It must be explicit, and  justified for simple devices (custom or \cots). The simplicity of a device relies on the simplicity and independence of all its functions, interfaces, building blocks, etc. The composition of simple items may therefore be a complex item.

\subsection{Considerations for accelerator-related objectives}
\label{sec:amc cots activities}

 
As per the \amcCOTS, most hybrid or multi-core architectures should fall under the definition of complex devices with multiple processing elements interacting. 
The Platform description and modelling phase for custom models, including any accelerator, will directly benefit from the \amcCOTS objectives' outcome, notably the conceptual and detail designs, and the device verification.
For \cots functions, as prescribed by the \amcCOTS objectives, a \pml model should be built from the manufacturer specification supplemented by characterisation and verification activities.
COTS IP specifically may provide detailed information on the function based on the stage of the design where they are instantiated, from Hard IP, embedded in the silicon by the manufacturer, to Soft ones, captured by a hardware description language.
Microcode, if present on used functions, needs to be considered as part of the platform model, 
as transactions between components.

We identified 4 activities for hybrid platforms and accelerators, per \amcCOTS objectives:
\noindent\textbf{Activity 1:}
An assessment should be performed for each device or its integration, as they may fall under different classifications: \cots, custom, soft IP, hard IP, \mcp... In particular, one should consider how the device is configured and accessed through hardware and software means, how it interacts with the rest of the system, and whether or not existing analysis techniques and tools apply.

\noindent\textbf{Activity 2:}
It is necessary to master \emph{complex} core architectures. 
More specifically stressing benchmarks would be needed in addition to documentation reviews.

\noindent\textbf{Activity 3:}
The utilisation of COTS must be within the limit of the device manufacturer specification.
This means that we need a specification of the COTS and its limits to check the compliance of usage.

\noindent\textbf{Activity 4:}
It is mandatory to qualify the  COTS behaviour and all micro-code, as defined in \amcCOTS (Section~\ref{sec:amc cots overview}).

\section{\amcMCP on hybrid architectures}
\label{sec:amc mcp}
The \amcMCP was extensively studied in \phylogun to define a certification methodology specifically for \multicore platforms~\cite{erts20}. We provide a brief summary of \amcMCP in the following.

\subsection{Overview of the \amcMCP}
\label{sec:amc mcp overview}

The \amcMCP defines a Multi-Core Processor (\mcp) as a device with two or more activated processing cores, with a core being a device that executes software. 
The \amcMCP recognises two exceptions to the definition of active cores, 
cores in lockstep executing the same software and inputs to compare their output;
  and cores connected solely through data buses typically used in avionics systems.

The AMC identifies both temporal and functional interference. Interference occur when the behaviour of an application varies over its behaviour in isolation when running in parallel with others. 
Interference occur as a result of shared hardware or software resources of the \mcp. 
As an example, interference may cause additional delays due to the arbitration of accesses to the resource or control flow variations due to external modifications of a shared variable.
Interference may cause a loss of deterministic behaviour for the application.

All software components should exhibit correct functional and timing behaviours in the presence of interference.
The AMC thus defines an interference channel as ``a platform property that may cause interference between software applications or tasks".
The impact of interference channels on applications in the system should be assessed.
The planning objectives in \amcMCP require the identification of shared resources, their use by, and their allocation to software applications, where applicable. 
This aims to first ensure the overall demand for resources at any given time does not exceed the available resources' capacity, and second to avoid or mitigate interference. 
Mitigations should be deployed and verified for impactful interference channels. 
The definition of an interference channel in the \phylog methodology is a conservative one, in line with the AMC objectives.

The objectives require all software hosted on the \mcp to be identified, including applications, operating systems, hypervisors, as well as libraries and runtime. 
The \amcMCP prescribes that any component for which interference is mitigated, possibly at the platform-level through robust partitioning, may be separately analysed and verified.
Otherwise, they should be tested on target with all other software components under the final configuration.
The \phylog methodology, and in particular interference calculus, can help assessing whether a modelled accelerator or a platform supports robust partitioning, by identifying interference channels, their impact, and that of any deployed mitigation (through benchmarking).

%

The question in the context of accelerators, is whether or not the \pml model is suitable to model them, and whether and how it should be extended.
Let us now characterize what type of resource is an accelerator.
We have identified 3 dimensions to take into account. 

\subsection{Dimension 1}
The first dimension concerns the number of applications that can simultaneously access  the accelerator.
We define two categories within that dimension:
\begin{itemize}
\item those that can be accessed  solely by one application at any given time are called \emph{unitary accelerators};
\item those that can be accessed by multiple applications simultaneously are called \emph{parallel accelerators}.
\end{itemize}

Note that the classification of an accelerator as \emph{unitary} may be inherent to the accelerator itself, e.g. if it cannot support multiple applications by design, or enforced by the platform, e.g. through application design or partitioning mechanisms.

\subsection{Dimension 2}

The second dimension concerns how
the accelerators are connected to the core and how the
workload is launched.
In that dimension, we have identified four categories.
The simplest case concerns \emph{tightly coupled} accelerators.
\begin{hybrid-case}
\label{case-coupled}
\textbf{Tightly coupled accelerator.}
The accelerator, as an example a vectorised functional unit, operates in the context of a complex core;
all transactions effectively originate from the core operations and transit through the core interfaces.

\noindent\textbf{Modelling impact on \pml.}
The core is still modelled as the sole initiator.
Such an accelerator can only be unitary, as a core executes only one application at any given time\footnote{\amcMCP explicitly excludes hyperthreading.}.
However transactions caused by an application using the accelerator may present a different profile.
\end{hybrid-case}

\begin{example}[of category \ref{case-coupled}]
The \armx A15~\cite{ArmA15TRM} cores can include a NEON \valu and floating point execution unit. 
SIMD Load/Store instructions
allow for transfers between NEON registers and the memory. 
Vector accesses target one or more lanes of the same or of consecutive vector registers.
The architecture thus does not guarantee the atomicity of the access to the memory even for scalar accesses. Each instruction can generate multiple transactions depending on the access size, the alignment of the address and the memory segment. 
Served by the private or shared caches, or the main memory, SIMD Load/Store may be subject to high timing variability and interference.

The A15 cores in the \keystone presented in Example \ref{ex-hybrid-keystone} do feature a NEON \valu. As discussed, the core is still modelled as a single initiator and the model in Figure \ref{fig:ex-keystone-pml} remains valid even when the NEON is in use.


\end{example}

The second case concerns \emph{passive} accelerators
that are controlled by a remote core, e.g. via configuration registers.
A passive accelerator cannot  generate any transaction to 
access any shared resource and
is thus a target that can be shared by several cores.
\begin{hybrid-case}
\label{case-passive}
\textbf{Passive accelerator.}
The accelerator is a resource used by the core(s).
It behaves from a high level point of view like a DDR that receives requests for \emph{load}
and \emph{store}.

\noindent\textbf{Modelling impact on \pml.}
It can be abstracted as a target. Two or more applications using the accelerator
concurrently would be assumed to interfere.
Thus it could be unitary or parallel, but in both cases it will be modelled in the same way.
The transactions caused by the controlling core may present a different profile.


\end{hybrid-case}

\begin{example}[of category \ref{case-passive}]
\label{ex-hybrid-nvdla-passive}

The NVIDIA Deep Learning Accelerator (NVDLA) outlined in Figure \ref{fig-hybrid-nvdla-passive} is an accelerator developed by NVIDIA, with both open-source hardware and software. The NVDLA is a complex \cots device. Tailored to neural network applications, it features functional blocks dedicated to convolution, activation functions, pooling, normalisation, or reshaping operations. The blocks can operate independently, performing memory-to-memory operations, or pipelined, passing data to each other to avoid the memory round-trip. The memory (DBBIF), interrupt (IRQ), and configuration (CSB) interface can be connected to various protocols such as \armx AXI. 

\begin{figure}[!htb]
    \centering
	\includegraphics[scale=0.3]{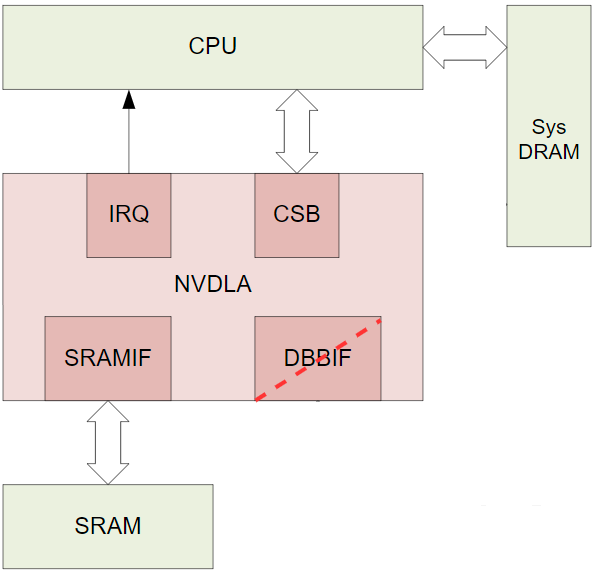}
	\caption{Integration of the NVIDIA NVDLA in a passive configuration~\cite{NVDLA}
	\label{fig-hybrid-nvdla-passive}}
\end{figure} 

As a soft IP, the NVDLA exposes all information regarding its internal behaviour which eases the development of a model for timing or interference analysis. The DBBIF, CSB, and target memory subsystem are obviously shared resources between functional blocks. The scope and mitigation of any resulting interference however require more information about the NVDLA integration. The device can be included as part of custom devices or available in future \cots platforms. 

\begin{figure}[!htb]
	\centering
     \includegraphics[scale=0.8]{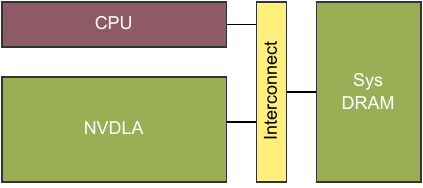}
     \caption{Simplified PML model for the NVDLA in a passive configuration}
	\label{fig-hybrid-nvdla-passive-pml}
\end{figure} 

Figure~\ref{fig-hybrid-nvdla-passive-pml} presents a PML model for a NVDLA in a passive configuration.
The accelerator and all its resources are abstracted as a single target, accessed through the interconnect.
Transactions initiated within the NVDLA remain within the device, e.g. from its functional blocks to the CSB or SRAM.
As such they would not need to be captured by the model.
They are thus implicitly assumed to be non-interfering with external transactions, e.g. from the CPU to the CSB.
Such an assumption must be verified during interference analysis.
\end{example}


The third case concerns \emph{semi-active} accelerators.
In that situation, the accelerator is triggered by a remote core but it
accesses shared resources (e.g. DDR) to \emph{load/store} its data.
Thus it
generates interferences within the hybrid architecture.
\begin{hybrid-case}
\label{case-semi-active}
\textbf{Semi-active accelerator.}
The accelerator operates under the control of a core
and it behaves from a high level point of view as a DMA that generates requests for \emph{load}
and \emph{store}
under the impulse of another core.
However the precise role of the core needs to be clarified, as well as the interface
between the accelerator and the hybrid platform.

\noindent\textbf{Modelling impact on \pml.}
A unitary \emph{semi-active} accelerator is thus modelled as a single initiator
and the profile of the remote core must contain all the transactions needed to configure the accelerator.
Parallel accelerators would need more refined analyses
to check whether they will be decomposed into one or multiple initiators.
\end{hybrid-case}

\begin{example}[of category \ref{case-semi-active}]
An example is the NVDLA in a "Small" configuration as depicted in Figure~\ref{fig-hybrid-nvdla-small}.
Compared to the passive configuration of Example~\ref{ex-hybrid-nvdla-passive}, the NVDLA accesses resources shared with other initiators in the system. 
 The NVDLA \cite{NVDLA} 
 could be modelled, as depicted in Figure~\ref{fig-hybrid-nvdla-small-pml}, using a single initiator with interfaces to the system,
as no interface or resource between the NVDLA and the controller is shared with other devices. 
This model assumes a pipelined configuration of execution on the NVDLA, where a single application may use the NVDLA and components do not interfere on the DBBIF.
(Example \ref{ex-hybrid-nvdla-large} considers a configuration where each functional block is a separate initiator.)
\label{ex-hybrid-nvdla-small}
\begin{figure}[!htb]
    \centering
	\includegraphics[scale=0.4]{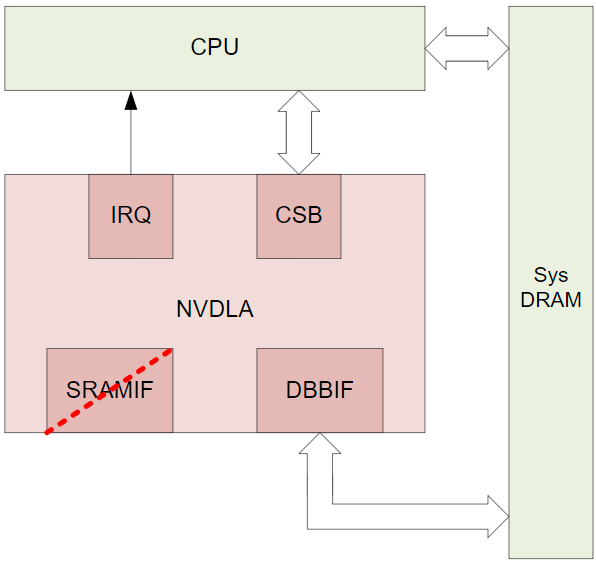}
	\caption{Integration of the NVIDIA NVDLA in a small configuration~\cite{NVDLA}
	\label{fig-hybrid-nvdla-small}}
\end{figure} 

\begin{figure}[!htb]
    \centering
     \includegraphics[scale=0.8]{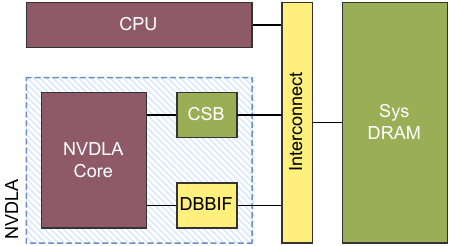}
     \caption{Simplified PML model for the small NVDLA}
	\label{fig-hybrid-nvdla-small-pml}
\end{figure} 
\end{example}

\begin{example}[of category \ref{case-semi-active}]	
\label{ex-hybrid-npu}
The i.MX 8M Plus processor from NXP \cite{NXPiMX8} 
features, amongst other accelerators, a NPU, e.g. a VIP8000 hard IP from VeriSilicon. The NPU is a complex \cots device. The processor reference manual unfortunately provides little information about the NPU, except for the high-level functional description in Figure \ref{fig-hybrid-npu}. It probably features \valu and systolic-like blocks as it supports hundreds of multiply and accumulate operations every cycle. The interface with the processor uses \armx AXI and  AHB bus interfaces which might help bound the demand of the NPU on the shared memory, and the interference it generates.

\begin{figure}[!htb]
	\centering
	\includegraphics[width=0.95\linewidth]{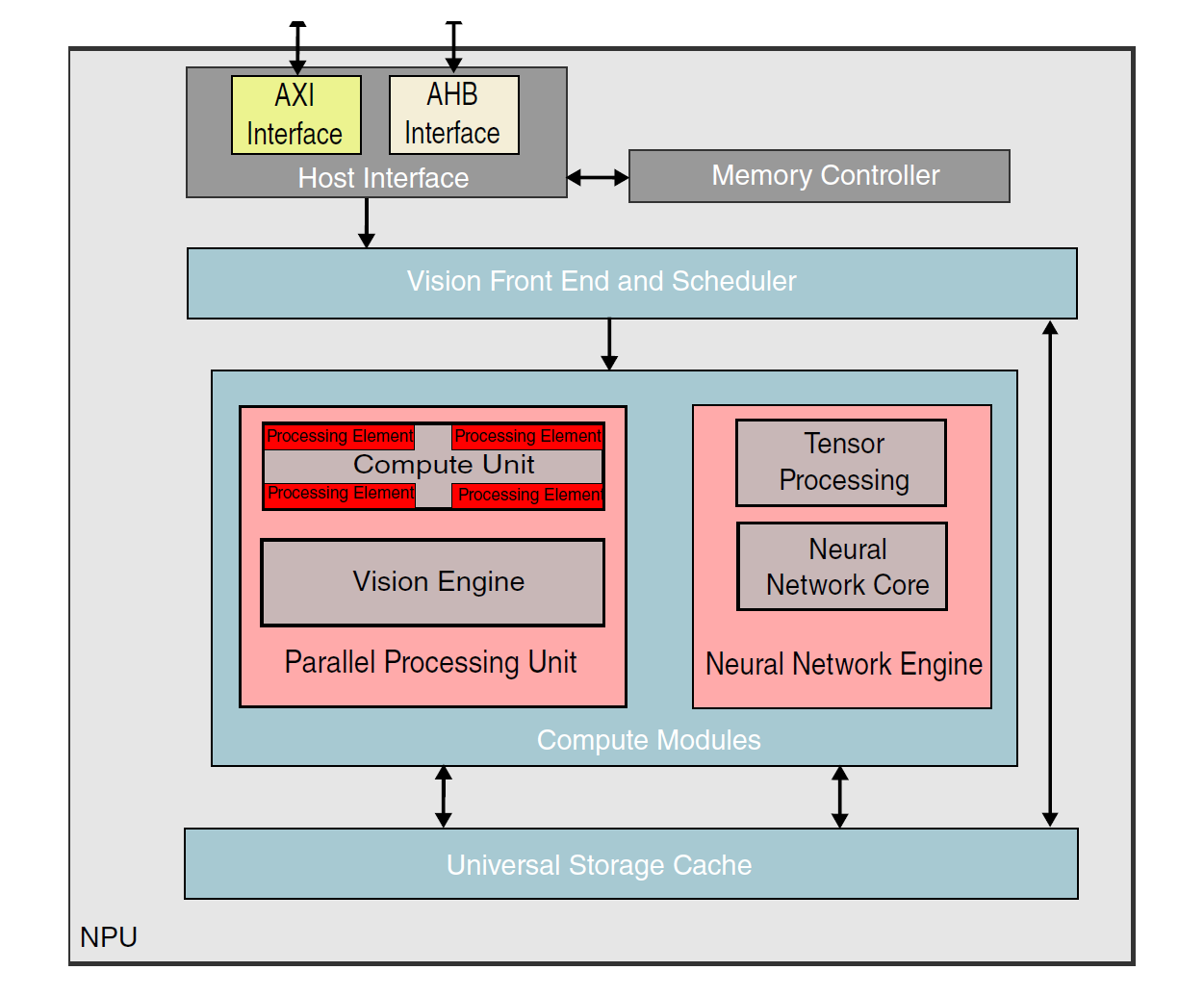}
	\caption{NPU High-level Block Diagram in the i.MX 8M Plus processor~\cite{NXPiMX8}
	\label{fig-hybrid-npu}}
\end{figure} 

It is difficult to model such a \cots device with no further information on its functional blocks, or without a characterisation by evaluation.
It could be abstracted as a single initiator.
This abstraction would need to be supported by limiting the use of the NPU as a \emph{unitary} accelerator, e.g. through platform configuration.
Furthermore, the abstraction will still require an assessment of the nature and volume of transactions the NPU generates.

\end{example}

The fourth case concerns \emph{active} accelerators.
An example of such accelerators are \gpu.
\begin{hybrid-case}
\label{case-active}
\textbf{Active accelerator.}
The accelerator operates independently and generates many \emph{load}
and \emph{store} transactions.

\noindent\textbf{Modelling impact on \pml.}
A unitary accelerator is thus modelled as a single initiator
where,
as for semi-active accelerators, parallel accelerators would need more refined analyses
to check whether they will be decomposed into one or more initiators.
\end{hybrid-case}

\begin{example}[of category \ref{case-active}]
When the accelerator is a \gpu used by a unique application at a time,
it can be modelled as an initiator and
single transaction forking to multiple targets should capture the combinations of behaviours of multiple threads running concurrently on the accelerator. Threads from the same application may not be considered as interfering with each other but with other applications in the system.
The \gpu scheduler decides upon execution of a computation kernel of the allocation of different blocks of threads to cores. The scheduling policy on most \cots platforms is subject to speculation, and the allocation of threads to cores is dynamic. 

In \pml, the initiator of a transaction from a given thread would thus be uncertain as well as for \amcMCP. Modelling the \gpu as a single initiator abstracts away this uncertainty. This should be a conservative, but sound abstraction for interference analysis between applications.
It needs to be backed by the platform to ensure only one task accesses the \gpu at any given time.
\end{example}

\begin{example}[of category \ref{case-active}]
When the \gpu is used simultaneously by several applications,
the \gpu cannot probably be modelled as a single unit. Different threads from different applications may share the \gpu cores, interfering on the \gpu internal resources and the shared platform resources. 
Uncertainty may arise in the mapping of threads to cores, and thus the generated interference by an application. 

\begin{figure}[!htb]
    \centering
 	\includegraphics[scale=0.7]{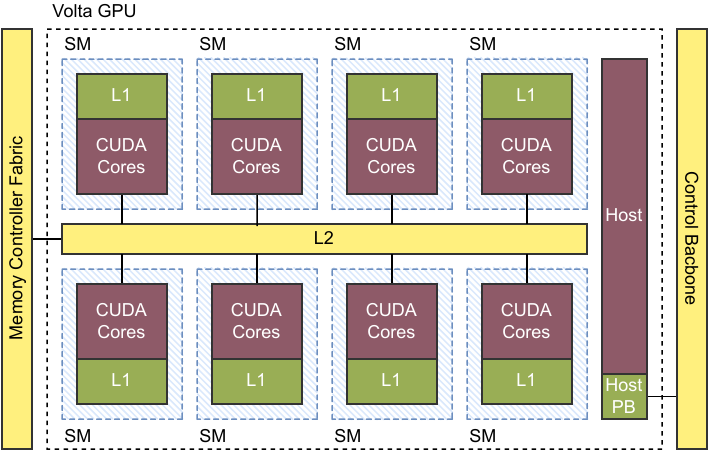}
	\caption{Simplified PML model for the Volta GPU}
	\label{fig-hybrid-volta-pml}
\end{figure}

However, the exact group of cores where an application is scheduled may not be relevant, provided said group is equivalent to the other groups of core on the platform. Capturing such platform symmetries in the \pml models would allow for some level of uncertainty. As illustrated in Figure~\ref{fig-hybrid-volta-pml}, \sm are symmetrical groups of cores on the Volta \gpu (Example \ref{ex-hybrid-volta}). Each \sm has the same number of cores and private resources.
Thus a group of threads should exhibit the same behaviour running in isolation in either \sm.
All \sm can access the same shared resources through the same paths on the Volta;
the interference suffered and generated by a group of threads is thus independent of 
the \sm where they run.
Isolating different applications to separate \sm does however rely on undocumented support from the platform~\cite{Bakita2023HardwareCP} (causing issues for Activity 3 in Section~\ref{sec:amc cots activities}).
\end{example}

\begin{example}[of category \ref{case-active}]
\label{ex-hybrid-nvdla-large}
\begin{figure*}[!tb]
    \centering
    \begin{minipage}{0.55\textwidth}
		\centering
		\includegraphics[scale=0.35]{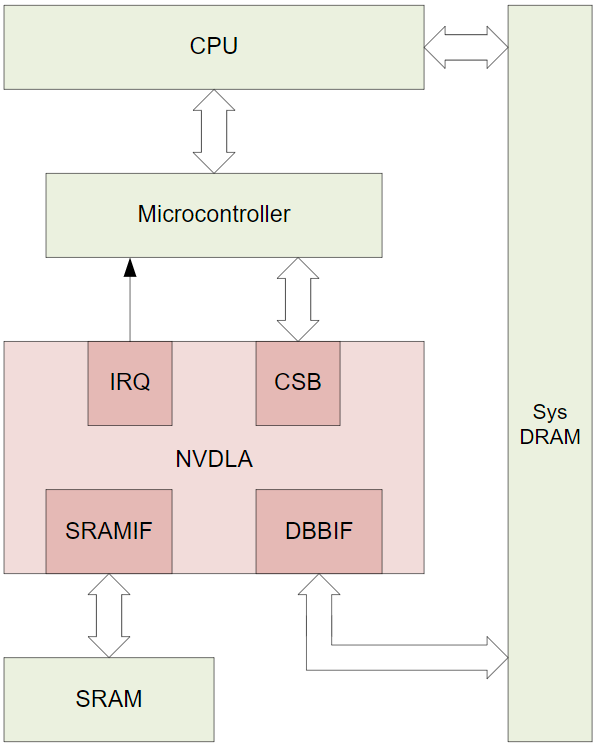}
		\caption{Integration of the NVIDIA NVDLA in a large configuration~\cite{NVDLA}
		\label{fig-hybrid-nvdla-large}}
    \end{minipage}\hfill
	\centering
    \begin{minipage}{0.45\textwidth}
	    \centering
	     \includegraphics[scale=0.70]{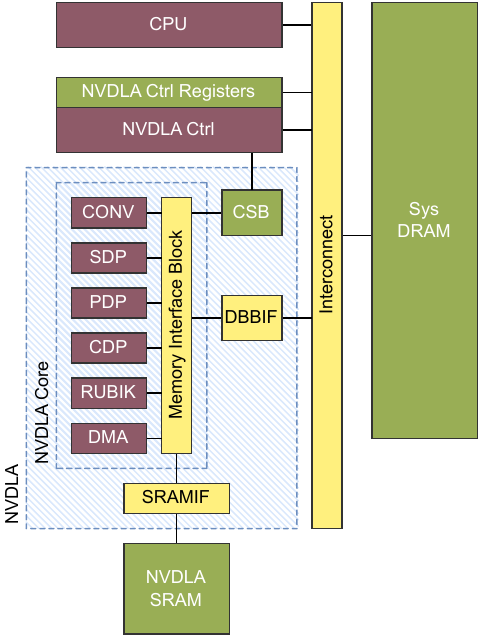}
	     \caption{Simplified PML model for the Large NVDLA}
		\label{fig-hybrid-nvdla-large-pml}
    \end{minipage}
\end{figure*} 
A NVDLA in a "Large" configuration features its own separate microcontroller, depicted in Figure~\ref{fig-hybrid-nvdla-large}, 
tightly coupled with the accelerator.
Where the CPU was in charge in the small configuration of Example~\ref{ex-hybrid-nvdla-small}, the microcontroller drives the accelerator.
Modelling the whole as a single accelerator would fail to distinguish transactions originating from the microcontroller and ones originating from the NVDLA functional blocks.
Each functional block of the NVDLA can be mapped to its own initiator, as depicted in Figure~\ref{fig-hybrid-nvdla-large-pml}.
This abstraction, compared to the one in Example~\ref{ex-hybrid-nvdla-small}, would allow transactions where one or more applications use the different functional blocks without interfering.
However each component (CONV, SDP, PDP...) may operate independently and interfere on the DBBIF. 

\end{example}



\begin{example}
The Xilinx ZYNQ-7000 AP \cite{ZynqSoC}, outlined in Figure \ref{fig-hybrid-zynq}, is a FPGA SoC with both Programmable Logic~(PL) and Processing System (PS). The PS features a 2-core A9 processor, with a NEON \valu, 
memory resources, and inputs/outputs. The processor offers multiple ports to connect PL devices to resources on the PS. Different ports may reach different or the same resources, through different protocols. Depending on if and how PL devices use said ports, the ports themselves or devices on the PL side may become shared resources and be classified as interference channels. 

\begin{figure}[!h]
	\centering
	\includegraphics[width=0.95\linewidth]{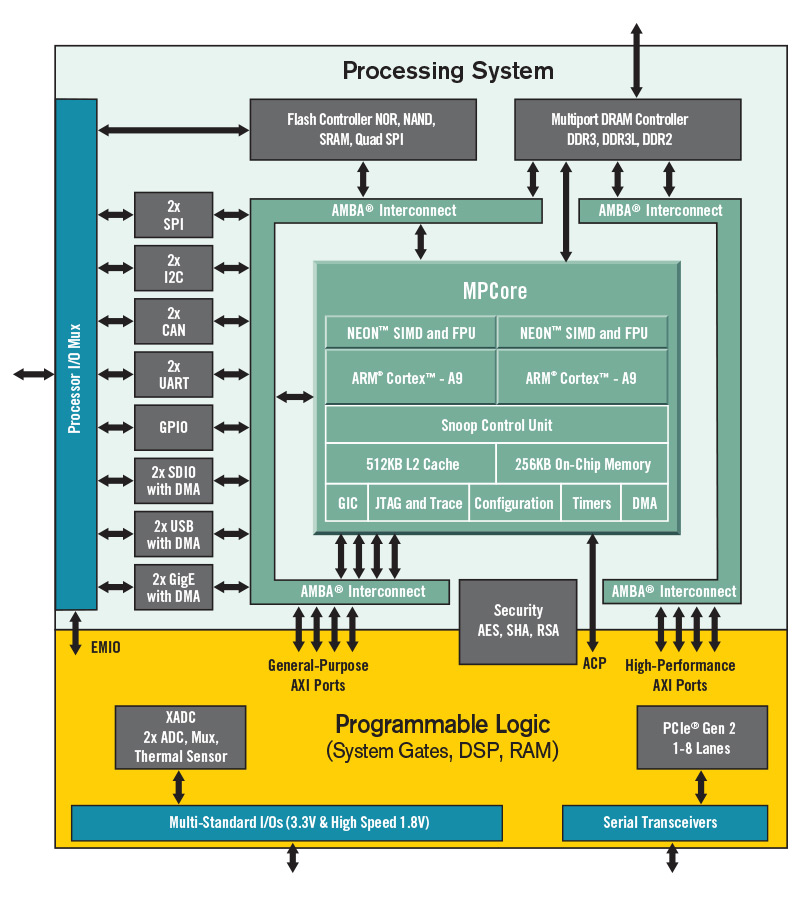}
	\caption{Overview of the Xilinx ZYNQ-A7000 AP
	\label{fig-hybrid-zynq}}
\end{figure} 

The PL features three types of ports: 4 general-purpose AXI ports (2 master and 2 slaves), 4 high-performance AXI master ports, and 1 AXI ACP port. The different ports first exhibit functional differences: as master ports cannot be used for the A9 processor to initiate reads from the PL. The AXI ACP port offers a high throughput and limited hardware coherency, as its accesses traverse the processor. However, it may result in serious cache trashing on the processor (as a result of invalidations), and interference on the A9 processor interconnect. The general-purpose ports allow access to most of the SoC interfaces, but share the interconnect with all input/output devices. The high performance ports only support high-throughput accesses from the PL to the main memory.

As a programmable logic device, the model for an \fpga is dependent on the devices and functions that have been configured, and on their use of the available platform resources.
As an example, a DMA configured on the PL may solely read memory from the flash controller using a general purpose port. 
It initiates transactions, contributing to and suffering from interference on shared resources. 
As such, it should be included as an initiator in the \pml platform model. 
Unless its interference is mitigated, it  should further be included as part of the final system configuration during analyses and tests. 

Care is thus required upon integrating devices on the PL side.
Each configured device should be considered and modelled per the aforementioned cases. 
The PS can be modelled as any platform. 
Existing interfaces to the PL or between the PS and PL should also be considered as part of the model most likely as transporters, based on their use by configured devices.
\end{example}

\subsection{Dimension 3}
\label{sec:amc mcp sw}

The third dimension concerns the  applicative layers that necessarily come with the accelerator,
e.g. a runtime used to offload work from the CPU to an accelerator. They
contribute to the interference generated on a platform. 
As an example the scheduling queue for a device may be shared between different applications, causing delays depending on the scheduler.
The transactions generated by an applicative layer also need to be characterised,
by assessing their documentation and their use of resources on the platform. 
The identification and verification must include all software running on accelerators as well as software interfaces or libraries used to program them. Some accelerators may indeed only be addressed through vendor-specific software interfaces.

\begin{example}
The definition and execution of kernels, functions running on the Volta \gpu, use the CUDA toolkit, or use higher-level libraries and runtimes which themselves offload computation on the GPU through CUDA. CUDA Kernels are written using a superset of a subset of C/C++. That is kernel code supports most of the C language, and the toolkit provides additional syntax for mapping code and data to the \gpu, or calling kernels. As such CUDA-enabled code cannot be analysed through existing tools as it may not parse as valid C/C++. 

As part of the CUDA toolkit are the compiler (\texttt{nvcc}) and assembler (\texttt{ptxas}) . The compiler is based on the mature LLVM compiler. The open-source nature of LLVM supports the verification of the generated code, and the development of compiler passes to support further analyses~\cite{7300224}. The assembler, which converts NVIDIA virtual assembly format into an executable binary, is closed. Information relevant for timing or coverage analysis may thus be lost at compilation.
\end{example}

\begin{example}
NEON instructions can be exploited through compiler optimisations, intrinsics, or assembly code.
Intrinsics are compiler- or vendor-provided functions often used to expose optimisations or vectorisation in languages without such constructs such as C. 
Compiler optimisations may jeopardize the traceability of the generated binary to the original source~\cite{7300224}, and \armx recommends the use of intrinsics over manual assembly code. Intrinsics explicit the use of vectorisation and of the NEON \valu. The added benefit is that the source code only exposes function calls, amenable to analysis.
\end{example}

\begin{example}
The software stack for the NVDLA comprises at its core the User-mode driver (UMD) and the Kernel-mode driver (KMD). The UMD loads a representation of a neural network, maps its inputs and outputs in memory, and informs the KMD that an inference job is ready. The KMD schedules available jobs, allocating DNN layers to function blocks, configuring the NVDLA registers, and collecting completed jobs. The KMD (and UMD) can run on the main CPU ("Small" system in Figure \ref{fig-hybrid-nvdla-small}) or through a dedicated core ("Large" system).

Similarly the open source software stack clearly identifies all required software, and opens the source code for analyses such as coverage or timing. Note that the NVDLA itself does not feature a core which executes user- or vendor-defined software. A NVDLA-enabled platform, depending on the integration, may not fall under the \mcp classification. Nonetheless, it still counts as one or more initiators as, once configured through the CSB, each block may initiate transactions to the memory. 
\end{example}

\begin{example}
The NPU is accessed through an OpenVX Driver. OpenVX \cite{OpenVX} is a standard and API which defines reusable computer vision and neural network functions. An OpenVX computation is expressed as a graph. Each node in the graph refers to its parameters and a kernel, the underlying  function. The standard defines a number of vision and neural network functions.
OpenVX is supported as a backend for numerous neural network runtimes through the Neural Network Runtime middleware \cite{iMXNN}. 

Nevertheless, the use of such runtimes raises several concerns. The transition from a model (computation graph) to software items is not explicit, and controlled by the runtime itself. This is not in line with the identification of software running on the platform as per \amcMCP. As the NPU supports only a subset of the OpenVX functions, runtimes may further elect to fallback to the CPU to run some software items. 
Using the NPU through the lower-level OpenVX driver would provide control over software items allocation between cores and the NPU. However, additional characterisation effort is still required to clarify the transactions the NPU might initiate.
\end{example}


\section{Related Work}
\label{sec:related work}



Worst-Case Execution Time (WCET) analysis methods~\cite{Wilhelm2008, wcet:ait, BallabrigaCRS10, hardy_et_al:OASIcs:2017:7303} rely on accurate processor models to produce conservative timing estimates of the execution of applications on a processor.
As such, the underlying processor models do often capture a more concrete and precise representation of the processor, e.g. accounting for the internal state of a core.
Those are finer-grained models than our transaction-based approach, but validating the underlying models may be a complex process~\cite{Sun2019}.
To the best of our knowledge, PasTiS~\cite{PastisWCETGPU} is one of the few efforts to build  a GPU model.

\pml takes inspiration from Initiator-Target modelling approaches found as an example in in~\cite{7778074}, where paths to shared resources are paramount to the interference analysis. The computation of interfering paths exponentially grows as a function of the number of initiators and targets. To cope with this issue, they propose to introduce reduction criteria (e.g., symmetries). 

(Memory) interference analysis approaches fall in two main categories: 
(1) Request-driven, which is based on a per-(memory) request analysis of an application~\cite{6925998}, 
(2) job-driven, which focuses on the number of (memory) requests of an application as a whole. 
Hybrid approaches blend the request-driven and job-driven~\cite{6925998}, i.e. considering both approaches jointly in a analysis~\cite{hassan_et_al:LIPIcs.ECRTS.2020.23}.

Model checking can be used to identify the interference of a platform as done in~\cite{nguyen:hal-02462085}. To do so, the approach uses 
formal languages for describing the behaviour of the application and multicore platform and introducing the interference concept and CADP toolbox to evaluate the model.

Interference mitigation techniques are used for minimizing, or even eliminating, the resource contention impact between processing cores. 
These techniques either make use of space (e.g., cache partitioning, bank parittioning) 
or time (e.g., scheduling, bandwidth reservation) partitioning to reduce the impact that interference entails. Survey~\cite{9714355} summarizes many of the techniques employed to this end. 


\section{Conclusion and perspectives}
\label{sec:conclusion}

We discussed the impact hybrid platforms on certification objectives for avionic systems.
Hybrid platforms embed several cores and accelerator devices in a small package, to provide high computational power while satisfying strict SWaP constraints.
We considered in particular two AMC: \amcCOTS for airborne electronic hardware, and \amcMCP for multi-core platforms.
Both require careful consideration about how devices are used and integrated in the system.

Most accelerators support highly parallel workloads and as such fall into the \amcCOTS \emph{complex} device category, and in scope of the \amcMCP.
As such, they require a thorough assessment of their behaviour and their integration in the platform.
We thus considered the use of \pml to capture and model knowledge about said devcices.
We identified 3 main dimensions relating to the hardware and software integration of the device in the platform, and proposed a related taxonomy.

We introduced a number of examples of COTS and Soft IP devices to illustrate the proposed taxonomy with \pml modelling templates.
COTS devices expose little information about their behaviour, and sometimes very limited control on said behaviour.
They thus require conservative assumptions and abstractions to comply with certification requirements.
Said abstractions have an impact on the performance of the accelerator and they do require backing by the platform configuration, e.g. a single \gpu user.

On the other hand, Soft IP (or custom devices), such as the NVDLA, do provide extensive information about their behaviour.
They also tend to offer higher configurability than COTS devices.
However, they do require separate objectives per \amcCOTS.
There might also be a vast amount of implementation and configuration choices to compare to select the most suitable integration w.r.t. to certification and performance objectives.

We did highlight that \pml is generic enough to model complex accelerators.
However, we also identified venues for improvements.
Accelerators such as \gpus cause uncertainty about the allocation of applications (threads) to initiators (cores), and thus the source of transactions.
The highly parallel nature of accelerators does also imply a high number of initiators in the system. This raises concerns about the required granularity of the platform model, the scalability of related analyses, and that of their output.

\section*{Acknowledgement}
\label{sec:acknowledgment}

The work presented in this paper is part of the PHYLOG 2 project supported by the Directorate General of Civil Aviation (DGAC). It is funded by the French government through the France Relance program, based on the funding from the European Union through the NextGenerationEU program.

The work presented in this paper has been funded in part by the Agence Nationale de la Recherche (ANR) under project ``ANR-22-CE92-0066-01''.

\bibliographystyle{IEEEtran}
\bibliography{bib}

\end{document}